\documentclass[10pt,a4paper,twopage]{scrartcl}
\usepackage[latin1]{inputenc}
\usepackage[english]{babel}

\usepackage{tikz}
\usepackage{cite}
\usepackage{amssymb,amsfonts,amsmath}
\usepackage{fancyhdr}

\usepackage[hang,small,bf]{caption}

\newcommand{\lc}{{\lambda_{\rm c}}}
\newcommand{\lt}{{\lambda_{\rm c}^{({\rm full})}}}
\newcommand{\lp}{{\lambda_{\rm c}^{({\rm monitored})}}}
\newcommand{\lr}{{\lambda_{\rm c}^{({\rm random})}}}
\newcommand{\taui}{{\tau_i}}
\newcommand{\tauj}{{\tau_j}}
\newcommand{\tauij}{{\tau_{(i,j)}}}
\newcommand{\fa}{{f_{\rm a}}}
\newcommand{\fl}{{f_\ell}}
\newcommand{\Pa}{{\Pi_{\rm a}}}
\newcommand{\Pl}{{\Pi_\ell}}
\newcommand{\Dt}{{\Delta t}}

\newcommand{\rpq}{{r_{p\to q}}}
\newcommand{\apq}{{a_{p\to q}}}

\begin{document}

\thispagestyle{empty} 

\begin{center}
 {\Large\bfseries\sffamily
 Impact of spatially constrained sampling of temporal contact networks on the evaluation of the epidemic risk\\
 }
 \vspace{1cm}\large
{Christian L. Vestergaard\textsuperscript{1,}*,
Eugenio Valdano\textsuperscript{2,}*,
Mathieu G\'enois\textsuperscript{1},
Chiara Poletto\textsuperscript{2},
Vittoria Colizza\textsuperscript{2,3}
and Alain Barrat\textsuperscript{1,3}}\\
\vspace{1.1cm}
{\footnotesize
\textsuperscript{1}Aix Marseille Universit\'{e}, Universit\'{e} de Toulon, CNRS, CPT, UMR 7332, 13288 Marseille, France\\
\textsuperscript{2}Sorbonne Universit\'es, UPMC Univ Paris 06, INSERM, Institut Pierre Louis d'\'Epid\'emiologie et de Sant\'e Publique (IPLESP~UMRS 1136), F75012, Paris, France\\
\textsuperscript{3}ISI Foundation, Torino, Italy.\\
*These authors contributed equally to this work.\\
}
\end{center}

 \vspace{2.2cm}

{\sffamily\footnotesize
\centering{\bfseries Abstract}\\
The ability to directly record human face-to-face interactions increasingly enables the
development of detailed data-driven models for the spread of directly transmitted infectious diseases at the scale of individuals.
Complete coverage of the contacts occurring in a population is however generally unattainable, due for instance to limited participation
rates or experimental constraints in spatial coverage.
Here, we study the impact of spatially constrained sampling on our ability to estimate the epidemic risk in a population
using such detailed data-driven models. The epidemic risk is quantified by the epidemic threshold of the 
susceptible-infectious-recovered-susceptible model for the propagation of communicable diseases, i.e. the critical value of 
disease transmissibility
above which the disease turns endemic. 
We verify for both synthetic and empirical data of human interactions that the use of incomplete data sets due to
spatial sampling leads to the underestimation of the epidemic risk.
The bias is however smaller than the one obtained by uniformly sampling the same fraction of contacts: it depends nonlinearly on the fraction of contacts
that are recorded and becomes negligible if this fraction is large enough. Moreover, it depends on the interplay between the timescales of population and spreading dynamics.
}

\newpage

\section{Introduction}

High-resolution, time resolved contact data describing face-to-face interactions in closed environments, such as hospitals, schools, conferences
or workplaces provide valuable information that can inform detailed models
of the spread of human airborne infectious diseases
\cite{Read2012,Salathe2010,Hashemian2010,Cattuto2010,Stehle2011a,Hornbeck2012,Gemmetto2014,Stopczynski2014,Obadia2015,Toth2015}.
In particular, wearable sensors enable the recording of
contacts with a spatial resolution of  $1$ to $2 \,{\rm m}$ and a temporal
resolution of the order of
seconds~\cite{Hui2005,ONeill2006,Eagle:2006,Salathe2010,Cattuto2010,Isella2011,Stehle2011a,Fournet2014,Toth2015}. However,
complete coverage of the contacts occurring within a population is generally unattainable.
As a result, the recorded network is usually a sample of the full underlying network of contacts,
and failure to take this into account may result in a biased assessment of
the vulnerability of the system to a spreading process~\cite{Ghani1998a,Genois2015b}.

Sampling is a well-known and well-studied issue, in particular in the context of static contact networks,
which are often collected by surveys or diaries. Various sampling procedures such as population sampling,
snowball sampling, or respondent-driven sampling, affect static
networks' measured properties in different ways, and many works have studied how
network characteristics such as the average degree, the degree distribution, clustering or assortativity depend
on the specific sampling procedure and on the sample size
\cite{Granovetter:1976,Frank:1978,Heckathorn:1997,Achlioptas:2005,Lee:2006,Kossinets:2006,Onnela:2012,Blagus:2015,Rocha:2016}. Other works have tackled the issue of
inferring network statistics from incomplete data \cite{Leskovec:2006,Viger:2007,Bliss:2014,Zhang:2015}.
Fewer studies have investigated how the outcome of simulations of dynamical processes in data-driven models is affected
if incomplete data are used, and few methods exist 
to obtain reliable estimates of the outcome of such processes when only 
sampled data are available  \cite{Ghani1998a,Ghani:1998b,Genois2015b}.

In the case of temporally resolved contact networks recorded using wearable sensors, two different sampling
effects are potentially present, leading to very different types of data loss.
First, limited rates of participation in the data collection campaign, with a
fraction of the population declining to wear sensors, lead to {\sl population sampling},
with the consequence that all contacts of non-participating individuals are absent from the data.
The use of such incomplete data in models of epidemic spread leads to an
underestimation of the epidemic risk, as the non-participating individuals are equivalent to immunized ones in simulations:
the absence of their contacts from the data removes potential transmission routes between the participating individuals.
Note that contacts with individuals that do not belong to the population
under study are also by definition absent from the data, but that this limitation may be less crucial if the population under scrutiny forms a coherent group.
Second, constraints stemming from the measuring infrastructure itself can represent another
source of data incompleteness: if contacts detected by the sensors
need to be uploaded in real time to radio receivers, the information corresponding to contacts taking
place outside the range of these receivers is lost \cite{Cattuto2010}.

Both types of sampling may affect data collection at the same time. As population sampling has been studied in~\cite{Genois2015b},  we focus here instead on the latter issue, which causes {\em spatially constrained sampling}. Such sampling leads to the absence of {\em some} of the contacts between participating individuals from the data set, namely 
those taking place outside of the monitored areas.
This sampling depends on the specific positions of the radio receivers
and on how individuals move in and out of the monitored locations.
The number of contacts each individual makes is thus underestimated in the data. Such sampling is thus also
expected to lead to an underestimation of the epidemic risk, in a way that
depends on the interplay between population dynamics and spreading dynamics.

To assess the impact of spatially constrained sampling on simulated spreading processes, we first consider an agent-based model of human interactions
that reproduces the phenomenology of empirical contact patterns observed in closed environments. We moreover assume that the agents can move between
two locations, similarly to individuals moving from one room to another. To mimic sampling, we consider the data obtained from the monitored location only, and compare it to the full data set of contacts taking place in the agents' population. 
We compute in both cases the epidemic risk as quantified by the epidemic threshold of the Susceptible-Infectious-Recovered-Susceptible (SIRS) model of infectious disease spread, of which the paradigmatic Susceptible-Infectious-Susceptible (SIS) and Susceptible-Infectious-Recovered (SIR) models are special cases. 
The epidemic threshold represents the critical value of disease transmissibility
above which the simulated pathogen is able to reach a large fraction of the population.
By comparing the values obtained for the partial and the full data sets,
we analyze the error made on the assessment of the system's epidemic risk when incomplete data is used.
To validate the results found for synthetic populations, we  next consider empirical face-to-face contact data collected at a scientific conference~\cite{Stehle2011a} and perform resampling experiments by selecting subsets of the full data composed of
the interactions taking place only in specific locations.

Our results show that the impact of spatially constrained sampling on the evaluation of the epidemic risk is qualitatively similar in
empirical and synthetic data.
First, the error on the epidemic threshold is much smaller than the one obtained from random sampling of the contacts.
Second, when the fraction of recorded contacts 
increases, the error decreases faster than linearly, until practically no error is made  above a certain fraction. We also observe some
discrepancies between the results obtained in real and synthetic contact networks and relate them to
the fact that individuals behave differently in different locations, an ingredient not present in the model used
to create the synthetic population and the contacts among its members.

The present paper is organized as follows.
In Section~\ref{sec:Threshold} we recall the definition of the Susceptible-Infectious-Recovered-Susceptible (SIRS) model of epidemic spreading, and we detail the computation of the epidemic threshold, 
 which is used to quantify the epidemic risk for a population.
In Section~\ref{sec:Model} we describe the model of human interactions that is used to generate synthetic data sets.
In Section~\ref{sec:Synthetic} we investigate the effect of spatial sampling on the estimate of the epidemic risk in a
synthetic population built with this model.
In Section~\ref{sec:Empirical} we consider an empirical network of face-to-face contacts, on which we perform spatially constrained resampling experiments, and compare the results with the ones obtained for synthetic data.

\begin{figure}
  \begin{tikzpicture}
    \node at(-3.4,5.3){ \includegraphics[width=0.35\textwidth]{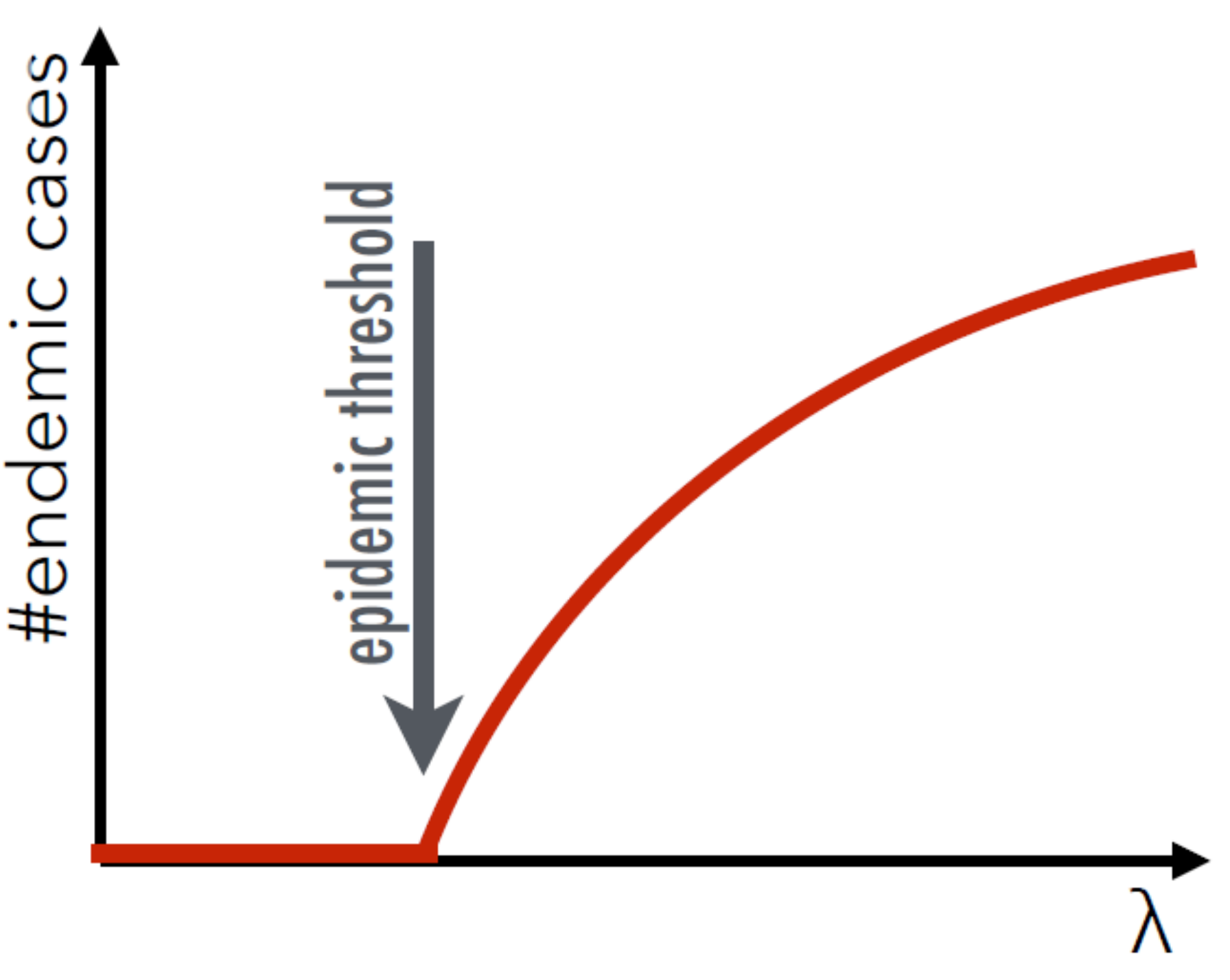} };
    \node at(-6.3,7){ {\large (a)}};
    \node at(3,5.1){ \includegraphics[width=0.45\textwidth]{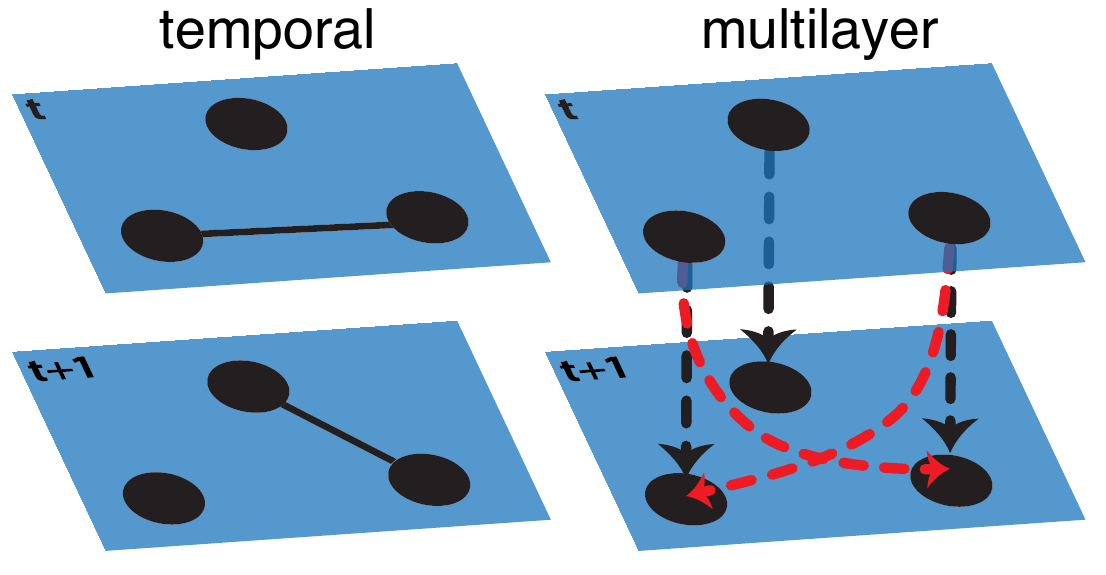} };
    \node at(-0.2,7){ {\large (b)}};

    \node at(-0.5,0){ \includegraphics[width=0.9\textwidth]{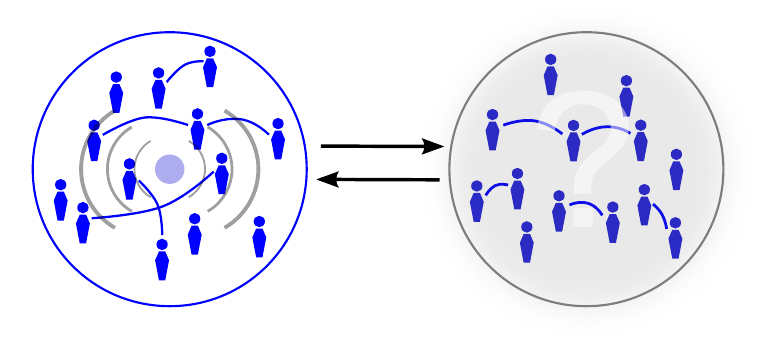} };
    \node at(-6.3,2.2){ {\large (c)}};
    \node at(-0.45,1.0){ \large $r_{1\rightarrow2}(\taui)$ };
    \node at(-0.45,-0.55){ \large $r_{2\rightarrow1}(\taui)$ };
    \node at(-3.8,-2.6){ \color{blue}{\large 1: monitored location}};
    \node at(2.85,-2.6){ \color{black}{\large 2: non-monitored location}};

    \node at(-0.5,-5.2){ \includegraphics[width=0.7\textwidth]{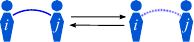} };
    \node at(1.45,-3.9){ \Large\textcolor{blue}{$\taui$} };
    \node at(3.9,-3.95){ \Large\textcolor{blue}{$\tauj$} };
    \node at(-3.7,-4.3){ \Large\textcolor{blue}{$\tauij$} };
    \node at(2.65,-4.3){ \Large\textcolor{blue}{$\tauij$} };
    \node at(-0.5,-4.6){ $\fl(\tauij)$ };
    \node at(-0.5,-5.9){ $\fa(\taui)\Pa(\tauj)\Pl(\tauij)$ };
    \node at(-3.7,-6.7){ \color{blue}{\Large Contact}};
    \node at(2.7,-6.7){ \color{blue}{\Large No contact}};
    \node at(-6.3,-3.6){ {\large (d)}};
  \end{tikzpicture}
  \caption{{\bf Illustration of the problem considered.}
  (a) We quantify the epidemic risk for a population by the epidemic threshold of the SIRS model, separating in the phase diagram a region in which the epidemic goes rapidly extinct from a region in which a finite fraction of the population is affected.
(b) Schematic representation of the multilayer mapping of a temporal network comprising $3$ nodes and $2$ time steps. The network on the left is mapped onto a 2-layer structure, with each layer containing a copy of all the nodes. 
Nodes are connected through directed links
to their future images  (black dashed) and to the future images of their present neighbors (red dashed).
  (c) To model the dynamics of spatially constrained sampling, we consider a population evolving in
 two separated locations; the full data set consists of all contacts taking place in both
locations, while the sampled data set considers only the contacts taking place in the ``monitored'' location.
Each agent $i$ moves between locations with rates $r_{p\to q}$ that depend on the time $\taui$ elapsed since
 she was last active, and can only have contacts with other agents present in the same location.
  (d) Rules governing interactions between agents within each location~\cite{Vestergaard2014}:
  the rate at which a contact between a pair of agents $(i,j)$ ends is controlled by the {\sl memory kernel} $\fl$ and depends on the time elapsed since the contact was created;
  the rate at which $i$ creates a contact with $j$ is controlled by the memory kernels $\fa$, $\Pa$, and $\Pl$,
which depend on the times elapsed since $i$ and $j$ either lost or gained a contact (respectively $\taui$ and $\tauj$)
and on the time $\tauij$ elapsed since $i$ and $j$ were last in contact.
  }
  \label{fig:problem}
\end{figure}
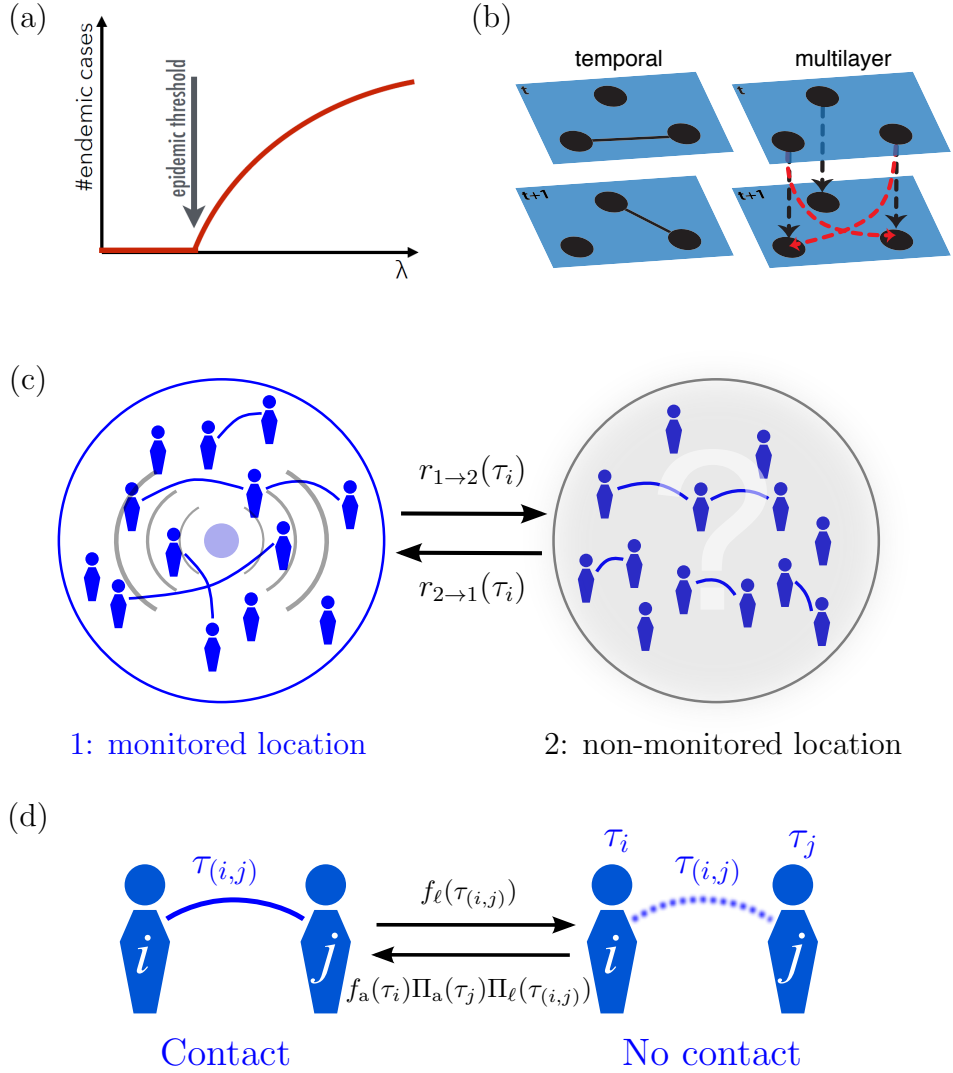

\section{Quantifying the spreading potential in a population---the epidemic threshold}
\label{sec:Threshold}

Let us consider a time-varying contact network~\cite{Holme2012} representing the temporally ordered sequence of contacts between individuals in a population:
individuals are represented by the nodes of the network, and at each point in time a link between two nodes indicates that the corresponding individuals are in contact.
In order to evaluate the vulnerability of the population to a disease that can spread through these contacts, we consider
the dynamics of the Susceptible-Infectious-Susceptible (SIS) and Susceptible-Infectious-Recovered-Susceptible (SIRS) models
on the contact network.
According to these models, an individual (agent) in the Susceptible (S) state, in contact with an agent in the Infectious (I) state, becomes infectious at rate $\lambda$. 
Infectious agents recover spontaneously at rate $\mu$, either going back to the Susceptible state (SIS model), or entering the Recovered state (SIRS model) where they are immunized to further infections.
In the SIRS model, the waning of immunity against the infection is modeled by letting recovered agents spontaneously enter the Susceptible state again at rate $\omega$.
At fixed rates of recovery $\mu$ and of loss of immunity $\omega$, the epidemic threshold $\lc$ is defined as the critical value of $\lambda$ that separates a regime where the epidemic rapidly goes extinct  ($\lambda<\lc$) from a regime where the disease becomes endemic ($\lambda>\lc$) [Fig.~\ref{fig:problem}(a)]~\cite{Pastor-Satorras2015}.
It can be found analytically for an arbitrary temporal network under
an individual-based mean field approximation, using the infection propagator approach introduced in~\cite{Valdano2015,Valdano2015b}. This method first introduces a mapping of the temporal network on a multi-layer network associating network's time frames to distinct layers. 
Within the framework considered here, the epidemic threshold is the same for the SIS and SIRS models, as well as for the SIR model (permanent immunity), which can be recovered as a special case of the SIRS model for $\omega=0$~\cite{Valdano2015b}.
For simplicity, we describe below the case of the SIS process and refer to~\cite{Valdano2015b} for the case of the full SIRS model. 

Assuming a generic temporal network of $N$ nodes evolving in discrete time, its evolution can be represented as a sequence of adjacency matrices $\left\{\mathbf{A}_t\right\}$, where $t=1,\cdots T$. $A_{t,ij}=0,1$ records the contact between nodes $i$ and $j$ at time step $t$. The SIS diffusion process on such network is shown to be equivalent to a new dynamic process unfolding on a particular multi-layer representation of the time evolution, described in Fig.~\ref{fig:problem}(b). By means of the supra-adjacency matrix formalism~\cite{DeDomenico2013,Cozzo2013,Wang2013} this new process can be formalized in terms of the following $NT\times NT$ block matrix, encoding both topology and spreading dynamics:
\[
\mathbf{M} =
\left(
\begin{smallmatrix}
 0                                             & 1-\mu+\lambda \mathbf{A}_1 &                                             0 &  \cdots  & 0 \\
 0                                             & 0                                             & 1-\mu+\lambda \mathbf{A}_2 &  \cdots  & 0 \\
 \vdots                                     & \vdots                                      & \vdots                                     &  \cdots  & \vdots \\
 0                                             & 0                                             &                                             0 &  \cdots  & 1-\mu+\lambda \mathbf{A}_{T-1} \\
 1-\mu+\lambda \mathbf{A}_T & 0                                             &                                             0 &  \cdots  & 0 \\
\end{smallmatrix}
\right).
\]
The resulting process consists in a diffusion on a static, albeit multilayer, network, for which $\lambda_c$ is known to be obey the relation $\rho[\mathbf{M}(\lambda_c,\mu)]=1$~\cite{Wang2003,Gomez2010}, where $\rho$ is the spectral radius of the matrix, i.e., the largest among the absolute values of the eigenvalues of the matrix. The computation can be further simplified by showing that this condition is equivalent to setting the spectral radius of another matrix equal to one, with the advantage that the latter is of size $N\times N$, thus not scaling with $T$. This matrix is the {\itshape infection propagator}:
\begin{equation}
\mathbf{P} = \prod_{t=1}^T \left(1-\mu+\lambda\mathbf{A}_t\right).
\end{equation}
 $\mathbf{P}$ encodes both network and disease dynamics, and its spectral properties fully characterize the epidemic threshold: $\rho[\mathbf{P}(\lambda_c,\mu)]=1$.

Given a data set represented by a temporal network of contacts, we will denote by $\lt$ the threshold computed using the full
temporal network. We will also consider subsets of contacts taking place in a specific location and subsets of contacts sampled uniformly at random. The resulting thresholds will be  denoted by
$\lp$ and $\lr$, respectively, and the impact of sampling will be measured by the ratios $\lt / \lp$ and $\lt / \lr$.

\section{Agent-based model of interaction dynamics}
\label{sec:Model}

In order to mimic spatial sampling,
we consider a population of $N$ agents who move between $2$ separate locations, and
can only interact with other agents present in the same location:
spatial sampling can indeed be simulated in a straightforward manner by considering that one of the locations is monitored, and the other is not, i.e., by excluding the contacts taking place in one of the locations from the data set [Fig.~\ref{fig:problem}(c)].
The rates of movements between the locations determine the fraction of sampled contacts.

We denote by $N_q(t)$ the number of agents in location $q(=1,2)$ at time $t$, where $N_1(t) + N_2(t)=N$.
The $N(N-1)/2$ pairs $(i,j)$ of agents are all potential {\em links}. If $i$ and $j$ are in contact the link $(i,j)$ is
{\em active}, while $(i,j)$ is {\em inactive} when $i$ and $j$ are not in contact.
At each time $t$, only the $N_1(t)(N_1(t)-1)/2 + N_2(t)(N_2(t)-1)/2 $
pairs of agents sharing the same location can have an active link.
Each agent $i$ is characterized by the time $\taui=t-t_i$ elapsed since the last time $t_i$ she changed state, i.e.,
the last time that she either gained or lost a contact or moved to a different location.
Links are characterized by their age, defined as the time $\tauij=t-t_{(i,j)}$ elapsed since the link was
either activated or inactivated [Fig.~\ref{fig:problem}(d)]~\cite{Vestergaard2014}.

We initialize the network with the agents randomly distributed in the different locations and all agents isolated (all links inactive).
We set $t_i=0$  for all agents and $t_{(i,j)}=0$ for all links.
The network evolves through the repetition of two sequential steps governing the agents' movements and contacts.
More precisely, at each time step $\Dt$,
\begin{enumerate}
\item  the locations of all agents are updated [Fig.~\ref{fig:problem}(c)]: Each isolated agent $i$ present in
location $p$ moves to location $q$ with probability $\Dt\,\rpq(\tau_i)$;

\item the contacts are updated [Fig.~\ref{fig:problem}(d)]:

(i) Each active link $(i,j)$ is inactivated (the contact between $i$ and $j$ stops) with probability $\Dt\,\fl(\tauij)$.

(ii) Each agent $i$ initiates a contact with another agent with probability $\Dt\,\fa(\taui)$.
The other agent $j$ is chosen among agents that are in the same location as $i$ and not in contact with $i$, with probability $\Pa(\tauj)\Pl(\tauij)$.
\end{enumerate}

These dynamical rules (only isolated agents can change location, and contacts can be initiated only
between agents in the same location) ensure that a link can be active only when the corresponding agents share the same
location. We note that
the model can easily be generalized to an arbitrary number
of locations. As such, it is  akin to metapopulation models composed by spatially referenced patches or subpopulations that are coupled together~\cite{Hanski1997, Grenfell1997, Tilman1997, Bascompte1998, Hanski2004}. These models generally assume homogeneous mixing within patches where the infection dynamics takes place (or mixing between population groups~\cite{apolloni_age-specific_2013}) and either an effective coupling between patches or 
explicit migration/mobility processes. While non-Markovian rules have been introduced in migration processes in metapopulation models for the study of disease spread and epidemic threshold conditions~\cite{keeling_individual_2010,balcan_phase_2011,belik_natural_2011,poletto_human_2013}, explicit contact structure between individuals in a patch have been rarely considered~\cite{Mata:2013}, assuming static topologies. Our approach thus differs from
usual metapopulation models in that it provides explicit temporally evolving contact structures within each group, allowing for different possible dynamics of mobility and contacts.

\begin{figure*}
  \begin{tikzpicture}
    \node at(0,0){ \includegraphics[width=0.7\textwidth]{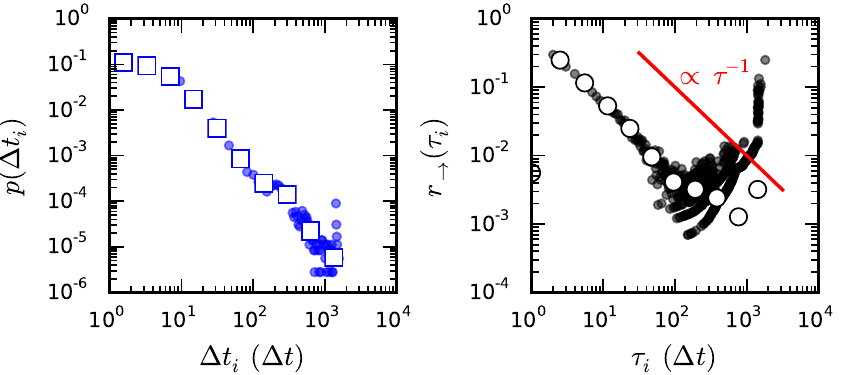}}; 
    \node at(-4.6,1.8){ \large(a) };
    \node at(.2,1.8){ \large(b) };
  \end{tikzpicture}
  \caption{{\bf Statistics of the movements of attendees at a scientific conference~\cite{Stehle2011a}.}
  (a) Distribution of the times $\Delta t_i$ individuals spent in a location that was monitored (visible) before leaving the visible area.
  (b) Rates $r_\rightarrow$ at which individuals left the monitored area as function of the time elapsed since they last either created a contact,
broke a contact, or arrived in the area, $\taui$.}
  \label{fig:movementRates}
\end{figure*}

The model's dynamics depends on the functional forms of the {\sl memory kernels} $r_{1\to2}$, $r_{2\to1}$ $\fl$, $\fa$, $\Pl$, and $\Pa$.
The kernel functions $\fl$, $\fa$, $\Pl$ and $\Pa$ measured from empirical
contact networks exhibit power-law like forms with exponents close to minus one \cite{Vestergaard2014}, indicating long term memory in the interactions.
Moreover, the movements in and out of monitored locations show similar long term memory (Fig.~\ref{fig:movementRates}), i.e., the rates $\rpq$ follow a similar power-law like shape with exponent approximately equal to minus one.
We therefore set $\rpq(\tau)=\apq(1+\tau)^{-1}$, $\fl(\tau)=z(1+\tau)^{-1}$, $\fa(\tau)=b(1+\tau)^{-1}$, $\Pl(\tau)\propto(1+\tau)^{-1}$,
and $\Pa(\tau)\propto(1+\tau)^{-1}$. Here $\Pa$ and $\Pl$ are normalized such that $\sum_{j \in q, j \notin V_i , j \ne i}\Pl(\tauij)\Pa(\tauj)=1$, where
the sum runs over all nodes $j \ne i$ in the same location $q$ as $i$ but not in contact with $i$
($V_i$ denotes the set of nodes in contact with $i$).

\section{Effect of dynamic sampling on the epidemic threshold}
\label{sec:Synthetic}

As discussed above,
the model emulates the sampling of empirical data caused by individuals moving in and out of monitored locations
[Fig.~\ref{fig:problem}(c)].
The parameters $N$, $b$, and $z$ are tuned such that the number of agents and the rates of creation and deletion of contacts are comparable to
those observed in empirical networks of face-to-face contacts~\cite{Vestergaard2014}. The parameters $a_{1\to2}$ and $a_{2\to1}$ control the
fraction of the total number contacts that occur in each location (Table~\ref{tab:parameters}). The total contact network
is composed by the contacts occurring in both locations, while spatial sampling is simulated by considering that only one of the locations
is monitored: the resulting sampled contact network is formed by the contacts taking place in the corresponding location only.
We calculate the epidemic threshold $\lt$ for the total contact network and  $\lp$ for the sampled one.
The discrepancy between the two is quantified by the ratio $\lt/\lp$, which is expected to be smaller than one as the sampled network underestimates the amount of contacts taking place in the population, in turn leading to an underestimation of the epidemic risk.

Note that we could in principle consider more than one monitored or non-monitored location, at the cost of additional parameters $a_{p\to q}$.
However, the important feature of the model for the problem at hand is its division into a (spatially separated) monitored part and non-monitored part.
An additional division of the monitored or non-monitored part into multiple subpopulations would only lead to a more complicated (if possibly more realistic) model of interaction dynamics in each part. We have actually
considered the case of three locations (one monitored and two non-monitored), finding 
qualitatively similar results to the case of two locations (not shown).
As we want to mimic empirical cases of few hundreds individuals in closed environments, we do not 
on the other hand consider the case of a large number of subpopulations as usually considered in metapopulation 
models of disease spread at the regional or global level.

\begin{table}
  \caption{{\bf Parameters and summary statistics of the synthetic data sets.}
  $f$: fraction of contacts that are recorded in the monitored location; $a_{1\to2}$, and $a_{2\to1}$: model parameters
fixing the rates of movement between locations;
  $N^*$: number of nodes that participate in at least one contact;
  $N^*_1$: number of nodes that participate in at least one recorded contact;
  $W$: cumulative duration of all recorded contacts;
  $N_{{\rm c}}$: total number of contacts recorded; $M_{1\rightarrow2}$ and $M_{2\rightarrow1}$: total number of movements from location 1 to 2 and 2 to 1, respectively.
  The total number of nodes in the networks is $N=450$ and $z=1.44$ in all cases.
For $f=12\%$ and $f=88\%$, $b=0.55$; for $f=20\%$ and $f=80\%$, $b=0.53$; for $f=39\%$ and $f=61\%$, $b=0.49$.
}
  \label{tab:parameters}
  \begin{tabular}{lcccccccc}
    \hline
    \hline
      $f$ & $a_{1\to2}$ &  $a_{2\to1}$ & $N^*$ & $N^*_1$  & $W$ & $N_{{\rm c}}$ & $M_{1\to2}$ & $M_{2\to1}$ \\
    \hline
    12\% & 2    & 0.15 & 384 & 364 & 10887 & 4924  & 7240  & 7248\\
    20\% & 1.3  & 0.25 & 394 & 383 & 22918 & 10914 & 15072 & 15097\\
    39\% & 0.5  & 0.3  & 406 & 400 & 47375 & 22864 & 16740 & 16805\\
    61\% & 0.3  & 0.5  & 406 & 404 & 74864 & 34167 & 16805 & 16740 \\
    80\% & 0.25 & 1.3  & 394 & 393 & 91631 & 40475 & 15097 & 15072\\
    88\% & 0.15 & 2    & 384 & 384 & 79247 & 34013 & 7248  & 7240 \\
    \hline
    \hline
  \end{tabular}
\end{table}

\begin{figure*}
  \centerline{
  \begin{tikzpicture}
    \node at(0,-0.6){ \includegraphics[width=1.1\textwidth]{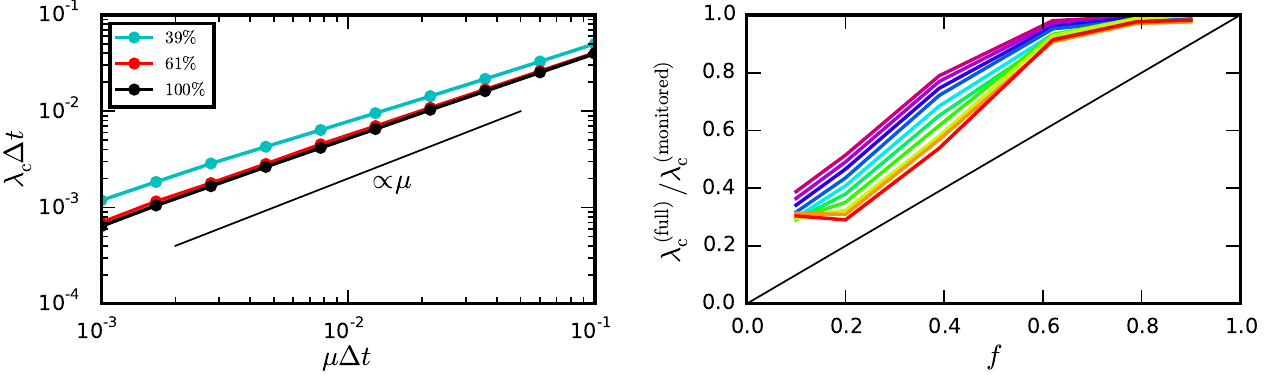} };
    \node at(-7.3,1.5){\large (a)};
    \node at(0.3,1.5){\large (b)};
    \node at(6.35,-0.55){ \includegraphics[height=.12\textheight]{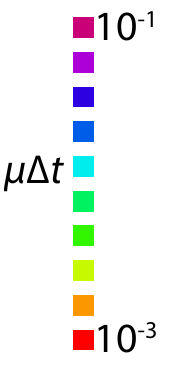} };
    \node at(-4.5,-5.7){ \includegraphics[width=0.4\textwidth]{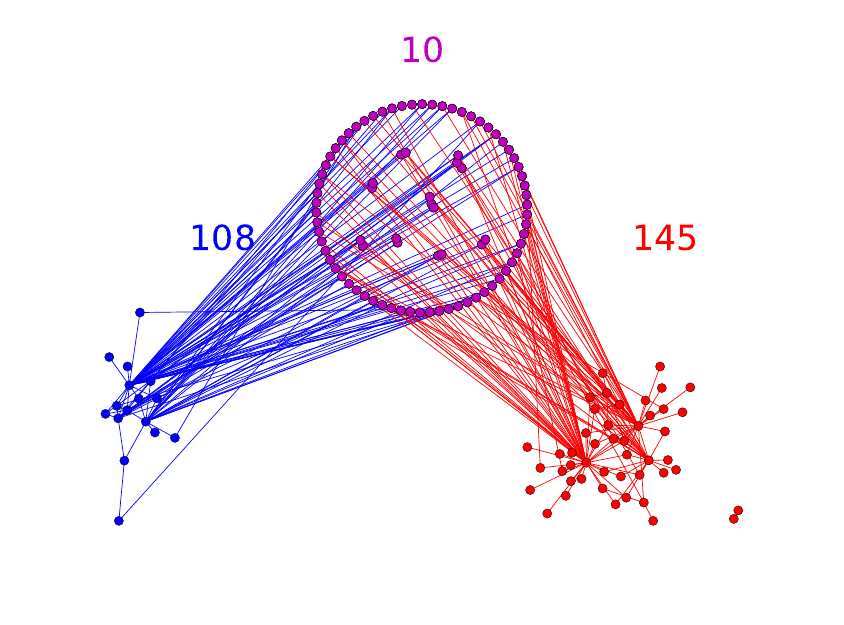} };
    \node at(0.5,-5.7){ \includegraphics[width=0.4\textwidth]{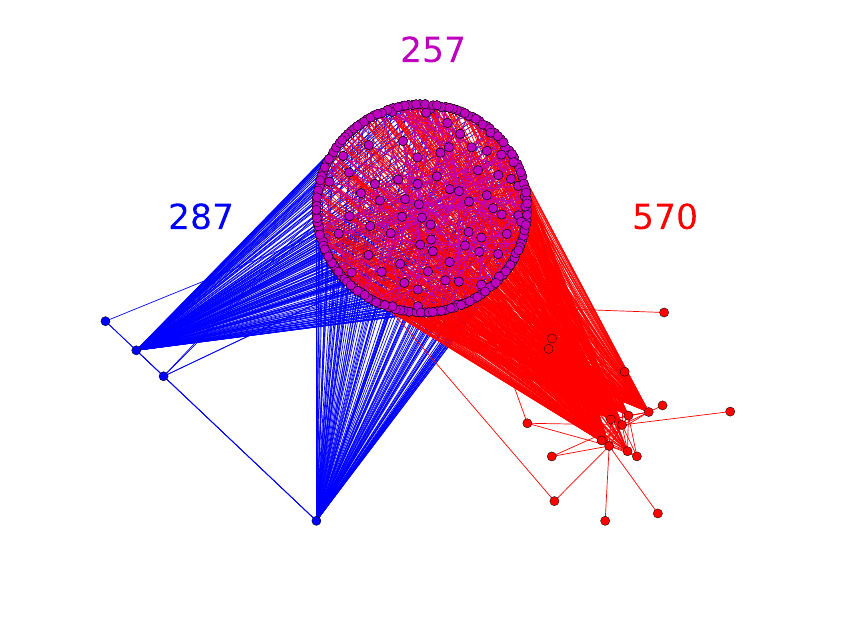} };
    \node at(5.5,-5.7){ \includegraphics[width=0.4\textwidth]{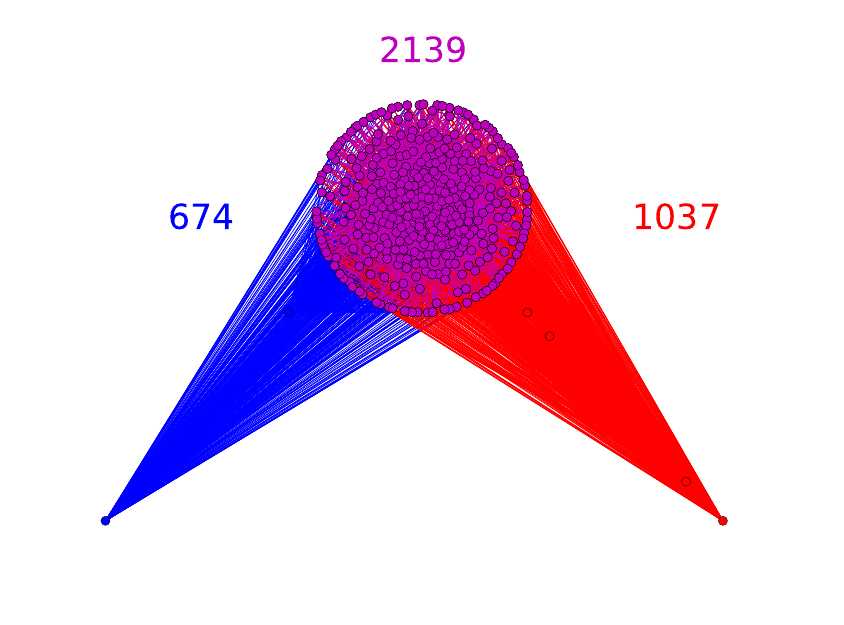} };
    \node at (-4.7,-3.3){\Large $10\,\Delta t$};
    \node at (0.3,-3.3){\Large $100\,\Delta t$};
    \node at (5.3,-3.3){\Large $1000\,\Delta t$};
    \node at(-7.3,-3.5){\large (c)};
    \node at(0,-10.2){ \includegraphics[width=1.1\textwidth]{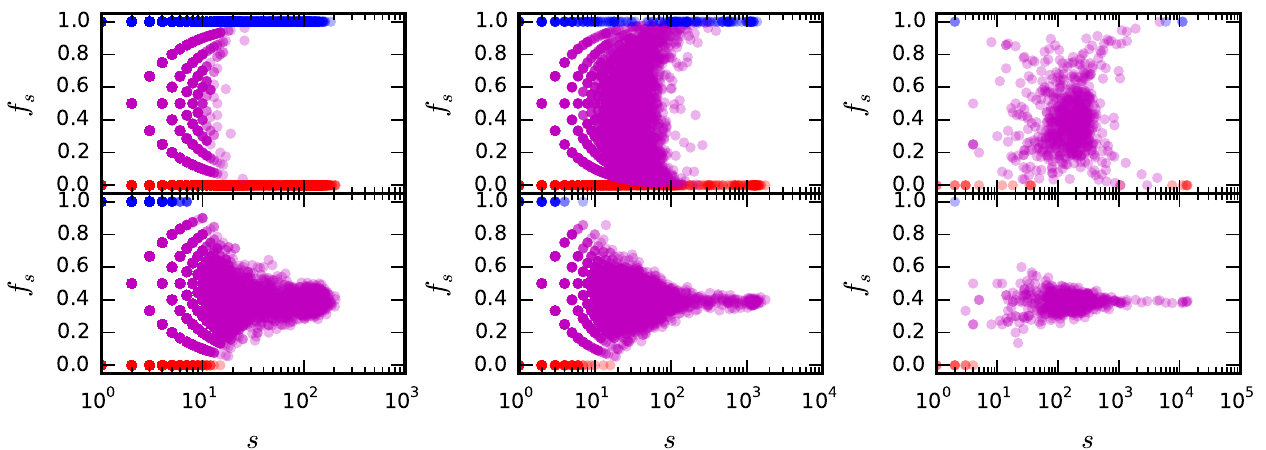} };
    \node at(-7.3,-8.){\large (d)};
  \end{tikzpicture}
  }
  \caption{{\bf (Color online) Effect of spatially constrained sampling on the epidemic threshold for synthetic networks.}
  (a) Epidemic thresholds as a function of $\mu$, calculated on model networks
where location $1$ ($39\%$ of the contacts), location $2$
($61\%$ of the contacts), or both locations ($100\%$ of the contacts) are monitored.
  (b) Ratio $\lt/\lp$ between the epidemic threshold calculated on the complete contact network ($100\%$) and on the monitored
part of the contact network, as a function of the fraction $f$ of the number of contacts that take place in the monitored location. The thin black line corresponds to the random sampling of contacts, $\lt/\lr=f$.
  (c) Examples of the contact network aggregated over $10\,\Dt$, $100\,\Dt$, or $1000\,\Dt$.
  Nodes in the network are divided into three groups: nodes for which all contacts are recorded in the sampled network (blue),
nodes for which part of their contacts are recorded (magenta), and
nodes for which no contacts are recorded (red). Numbers give the number of
links for which all (blue), part of (magenta), or no (red) contacts are recorded in the sampled data.
 (d) Fraction $f_s$ of contacts that are recorded over the given time-scale as above for nodes
that have at least one contact in the monitored location, versus strength of the node in the complete network
(the strength of a node is given by the sum of the durations of its contacts). Top plots correspond to
a spatial sampling while bottom plots correspond to random sampling of contacts; colors correspond to the groups of (c).
  (c),(d) show the case where $39\%$ of the contacts are recorded.
  Simulations of network dynamics were performed for $1000\,\Dt$ before recording contacts in order to ensure that the system had reached a quasi-stationary state. 
}
  \label{fig:modelResults}
\end{figure*}
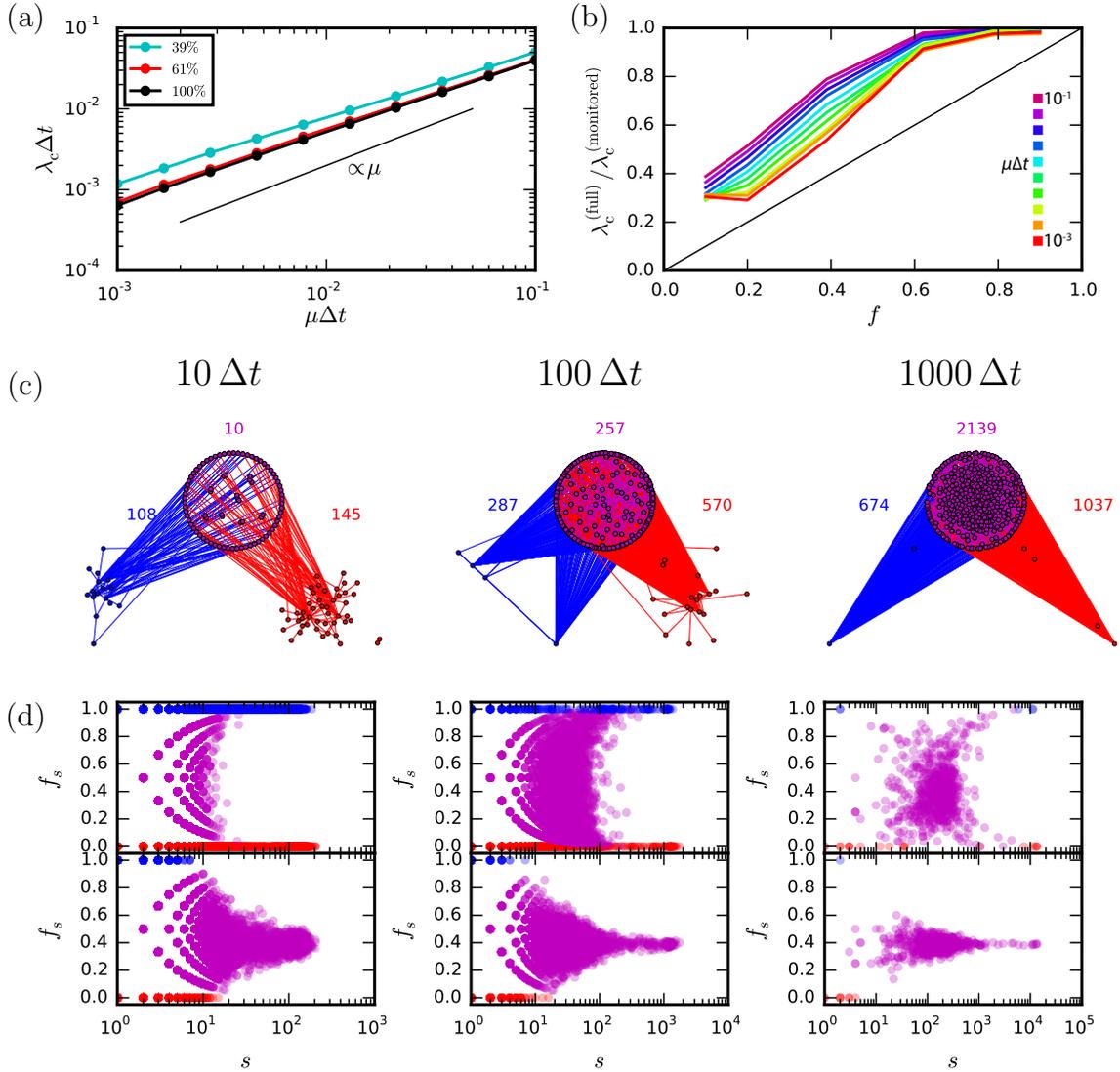

Due to the heterogenous nature of the network dynamics, the thresholds $\lt$ and $\lp$ increase sub-linearly with $\mu$
[Fig.~\ref{fig:modelResults}(a)]. This is explained by the presence of temporal
correlations leading to repetition of contacts in local groups of connected individuals ({\em ``temporal cliques''}), which facilitate
the persistence of the disease, thus decreasing $\lc$~\cite{Sun2015}. This decrease is larger for larger $\mu$ (faster timescales)
as the spread on long timescales is less sensitive to temporal patterns \cite{Stehle2011a}.
This effect is also slightly stronger for the sampled networks than for the full network due to temporal
cliques naturally being localized in a single location, and not spanning the two.

Figure \ref{fig:modelResults} summarizes our main results. First, we note that
the estimate $\lp$ obtained from spatially constrained sampling of the contacts (due to the monitoring of only one
of the two locations) is much closer to the true threshold $\lt$
than {what we would obtain if contacts were simply
sampled at random [Fig.~\ref{fig:modelResults}(b)]: for random sampling of a fraction $f$ of the contacts, we would indeed
find $\lt/\lr=f$ (thin black line).
Moreover,  $\lp$ is closest to $\lt$ for high values of $\mu$, corresponding to fast spreading.
We can understand this effect as due to the interplay of the timescale of the spreading procees, set by $1/\mu$,
with the timescales of the nodes' movements and contacts.
For fast processes, fewer individuals change location over the relevant time-scale $1/\mu$. As a result,
most links are either completely recorded or not present at all in the sampled data [Fig.~\ref{fig:modelResults}(c), left panel].
Moreover, as shown in Fig.~\ref{fig:modelResults}(d), nodes with a high strength (the {\sl hubs}), which
have a crucial effect on the epidemic threshold~\cite{Boguna2013}, tend to have all of their contacts in a
single location on such short time-scales, in contrast with the case of random sampling of an equivalent amount of contacts.
Hence, the heterogeneous character of weights and strengths is better conserved than for random sampling.
On long timescales on the other hand, most links are recorded partially [see right panel of Fig.~\ref{fig:modelResults}(c)]
and for most nodes, even hubs, only part of their contacts are monitored [Fig.~\ref{fig:modelResults}(d), right panel]: this
makes the resulting networks more similar to the random case. Note however that even for long time-scales, the distribution of the measured fraction of
nodes' contacts remains heterogeneous and differs significantly from the distribution in the case of random sampling of contacts, where the
distribution is localized around the fraction $f$ of sampled contacts  [Fig.~\ref{fig:modelResults}(d), right panel].
As a result, $\lp$ remains a markedly better estimate of $\lt$ than $\lr$, even for slow infection dynamics.
This is due to the scale free nature of the movement dynamics, which implies that time-scales larger than $1/\mu$ are always
represented in the dynamics. If instead we
consider a model where individuals move between locations according to a
Poisson process with constant rates $r_{1\rightarrow2}(\tau)=a_{1\rightarrow2}$ and $r_{2\rightarrow1}(\tau)=a_{2\rightarrow1}$, we find that $\lp$ approaches $\lr$ for small $\mu$,
as shown in Fig.~\ref{fig:PoissonModelResults}.

\begin{figure*}
  \centerline{
  \begin{tikzpicture}
    \node at(0,0){    \includegraphics[width=0.55\textwidth]{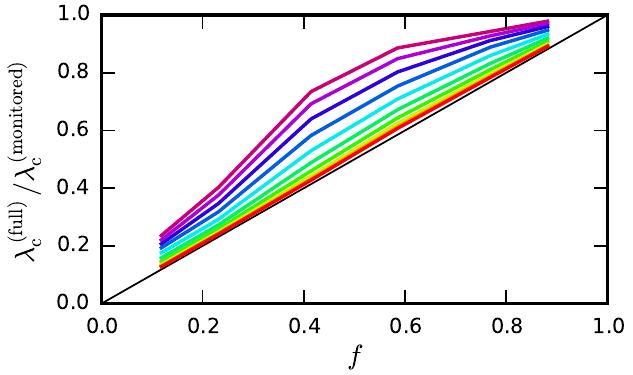} };
    \node at(2.65,0.05){ \includegraphics[height=.12\textheight]{legend.pdf} };
  \end{tikzpicture}}
  \caption{{\bf (Color online) Effect on the epidemic threshold of dynamic sampling of model with Poissonian movement dynamics.}
  Ratio $\lt/\lp$ between the epidemic thresholds calculated on the complete contact network and on the monitored part
of the contact network, as a function of the fraction $f$ of the number of contacts that take place in the monitored location. The thin black line marks the result for random sampling of contacts, $\lt/\lr=f$.
Parameters are chosen such that summary statistics are similar those of the model with scale-free movements (Table 1).
  Simulations of network dynamics were performed for $1000\,\Dt$ before recording contacts to ensure that the system had reached a quasi-stationary state.  }
  \label{fig:PoissonModelResults}
\end{figure*}

\section{Empirical data}
\label{sec:Empirical}

To validate the results found for model networks, we consider an empirical temporal network of face-to-face contacts measured at a scientific conference by the SocioPatterns collaboration (www.sociopatterns.org).
As described in \cite{Stehle2011a},
participants in the two-day conference were asked to wear RFID tags (see \cite{Cattuto2010}) tuned in order to register
close face-to-face proximity (1 to 2 m). Such contact events detected by the tags were immediately sent to a number
of receivers installed in the environment.
The conference took place in a large building with several separate areas, three of which were monitored by $12$ radio receivers: $5$
receivers were placed in a room called ``Rhodes'', $4$ in another room called ``Muses'' (both rooms were used as exhibition halls where
many contacts occurred), and $3$ in the entrance hall of the building (Supplementary Fig.~\ref{fig:Map}).
We resample the data set by
dividing it into subsets composed of the contacts recorded by different groups of {receivers}~(see Table~\ref{tab:Empirical}),
which we refer to as {\sl locations}.
We then compare the epidemic threshold computed using the full data set to the ones obtained from each such subset.
In order to check the effect of the finite data set length, we have moreover proceeded as in \cite{Valdano2015}:
we have computed $\lt$, $\lp$ and their ratio for increasingly larger values of the period $T$
(see section \ref{sec:Threshold}) up to the entire data-collection time window. We have observed
a convergence of all three quantities for $T$ larger than half of the data temporal length,
indicating that the data-collection period is long enough to characterize the epidemic dynamics.

We note that the situation is not completely analogous to the model described in the previous sections, which accounts for synthetic dynamics of movements of individuals between two locations only. 
Here, individuals move between more than two different locations (the two monitored rooms, the hall, and the locations in the building that were
out of the range of the receivers), 
their movement and interaction dynamics are non-stationary,
and their interaction behavior, as we will see, differs between different locations. 

\begin{table}
  \caption{{\bf Summary statistics for the empirical data.}
  Location: subset number;
  Receivers: receivers included in the subset;
  $f$: fraction of contacts that are recorded in the monitored location; $b$, $N^*_1$: number of nodes that participate in at least one recorded contact;
  $W$: cumulative duration of all recorded contacts;
  $N_{{\rm c}}$: total number of contacts recorded;
  $M_{\to1}$ and $M_{1\to}$: total number of movements to and from the monitored location, respectively.}
  \label{tab:Empirical}
  \begin{tabular}{llcccccc}
    \hline
    \hline
      Location & Receivers & $f$ & $N^*_1$ & $W$ & $N_{{\rm c}}$ & $M_{1\to}$ & $M_{\to1}$ \\
    \hline
     1 (Room ``Muses'') & \{101, 108--110\}  & 9\% & 306 & 5209 & 2454 & 3083 & 3085\\
     2 (Hall) & \{100, 103, 111\}  & 14\% & 365 & 8408 & 3912 & 3421 & 3413\\
     3 & \{100, 101, 103, & 22\% & 387 & 13609 & 6352 & 6763 & 6759\\
       & \qquad\ \ 108--111\}\\
     4 & 107  & 44\% & 380 & 26781 & 10846 & 10864 & 10861\\
     5 (Room ``Rhodes'') & \{102, 104--107\} & 79\% & 393 & 48440 & 17603 & 14276 & 14277\\
     6 & \{100, 102--107  & 91\% & 403 & 55925 & 21834 & 19717 & 19707\\
       & \qquad\qquad\ \  111\} \\
     7 & \{100--106,  & 93\% & 403 & 56819 & 21455 & 17578 & 17577\\
       & \qquad\ \ 108--111\} \\
     All & \{100--111\} & 100\% & 403 &  61242 & 23279 & 19616 & 19616\\
    \hline
    \hline
  \end{tabular}
\end{table}

Notwithstanding, we observe the same overall behavior obtained for the synthetic data sets, as shown in Fig.~\ref{fig:empiricalResults}.
First, both $\lt$ and $\lp$ depend non-linearly on $\mu$ [Fig.~\ref{fig:empiricalResults}(a)]. Second,
$\lp$ is a much better estimate of $\lt$ than $\lr$, and the error made when using the resampled data
becomes negligible for large enough fractions of observed contacts [Fig.~\ref{fig:empiricalResults}(b)].
Figure \ref{fig:empiricalResults}(b) also shows an interesting
qualitative difference between the results obtained for the resampled empirical data and the synthetic data sets:
while  $\lt/\lp$ always increases with $\mu$ for the synthetic data sets (better estimation for faster processes, as discussed above),
this is not always the case for the empirical data.
Consider for example locations $3$ and $4$, corresponding to $f=22\%$ and $f=44\%$
[Fig.~\ref{fig:empiricalResults}(b)].
For location $3$, the estimate of the threshold is more accurate for faster processes (larger $\mu$), similarly to the results of
Fig.~\ref{fig:modelResults}(b).
Conversely, for location $4$, the estimate is more accurate for slower processes, with $\lt/\lp$ very close to $1$ for small enough $\mu$.

\begin{figure*}
  \centerline{
  \begin{tikzpicture}
    \node at(0,-0.6){ \includegraphics[width=1.1\textwidth]{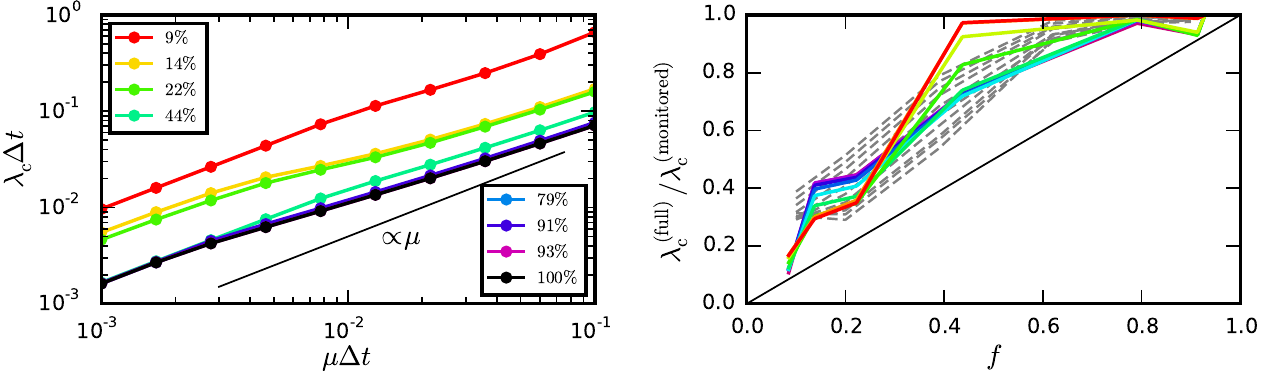} };
    \node at(-7.3,1.5){\large (a)};
    \node at(0.3,1.5){\large (b)};
    \node at(6.35,-0.55){ \includegraphics[height=.12\textheight]{legend.pdf} };
  \end{tikzpicture}
  }
  \caption{{\bf Effect on the epidemic threshold of spatial resampling of an empirical contact network.}
  (a) Epidemic thresholds as function of $\mu$, for the different locations corresponding to the fractions of the total number of contacts listed in the legend.
The curves corresponding to $f=91\%$ and $f=93\%$ are almost superimposed underneath the black one ($f=100\%$).
  (b) Ratio $\lt/\lp$ between epidemic thresholds calculated on the complete and partial empirical contact networks (full colored lines). The ratios
$\lt/\lp$ obtained for the synthetic networks (section \ref{sec:Synthetic}) are shown as grey dashed lines for reference.}
  \label{fig:empiricalResults}
\end{figure*}

We argue that this discrepancy between results obtained with synthetic and empirical data sets
is due to structural differences between real locations.
In the model, we impose the same microscopic dynamics for contact formation and deletion in both locations,
while this may not be the case in the empirical data set; individuals may behave differently in different locations,
leading to different contact patterns.

Figure~\ref{fig:plot_strength_behavior}(a)-(b) confirms this picture by investigating the contribution to the strength of each node of the
contacts taking place in locations $3$ and $4$, as a function of the node's strength rank in the full data set. The comparison of the empirical case
with the result of a random sampling of contacts shows that the
 hubs (nodes with the highest strength) are significantly over-represented with respect to the random case
in location $4$, while they are under-represented in location $3$. For slow spreading diseases,
using the weighted aggregated network in simulations yields a good approximation of the outcome of processes on
the complete temporal network~\cite{Stehle2011a,Valdano2015b}. Therefore, when using the data collected in location $4$,
we obtain a particularly accurate estimate of the threshold for slow processes because the hubs of the full network
have most of their activity precisely in this location.

Figure~\ref{fig:plot_strength_behavior}(c)-(d) moreover shows that, in the case of the synthetic data sets,
no systematic under- or over-representation of the hubs of the full network is observed in the monitored location,
as expected since locations $1$ and $2$ are equivalent in terms of contact dynamics in the model.

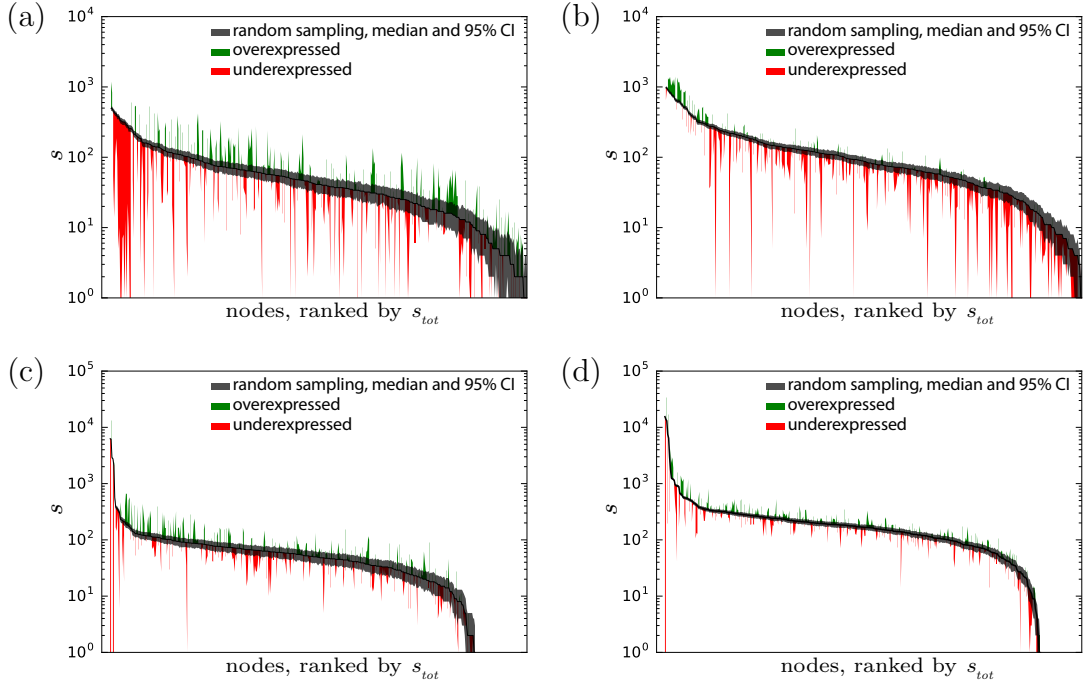
\begin{figure}
  \centerline{
  \begin{tikzpicture}
    \node at(-3.3,-0.3){ \includegraphics[width=0.55\textwidth]{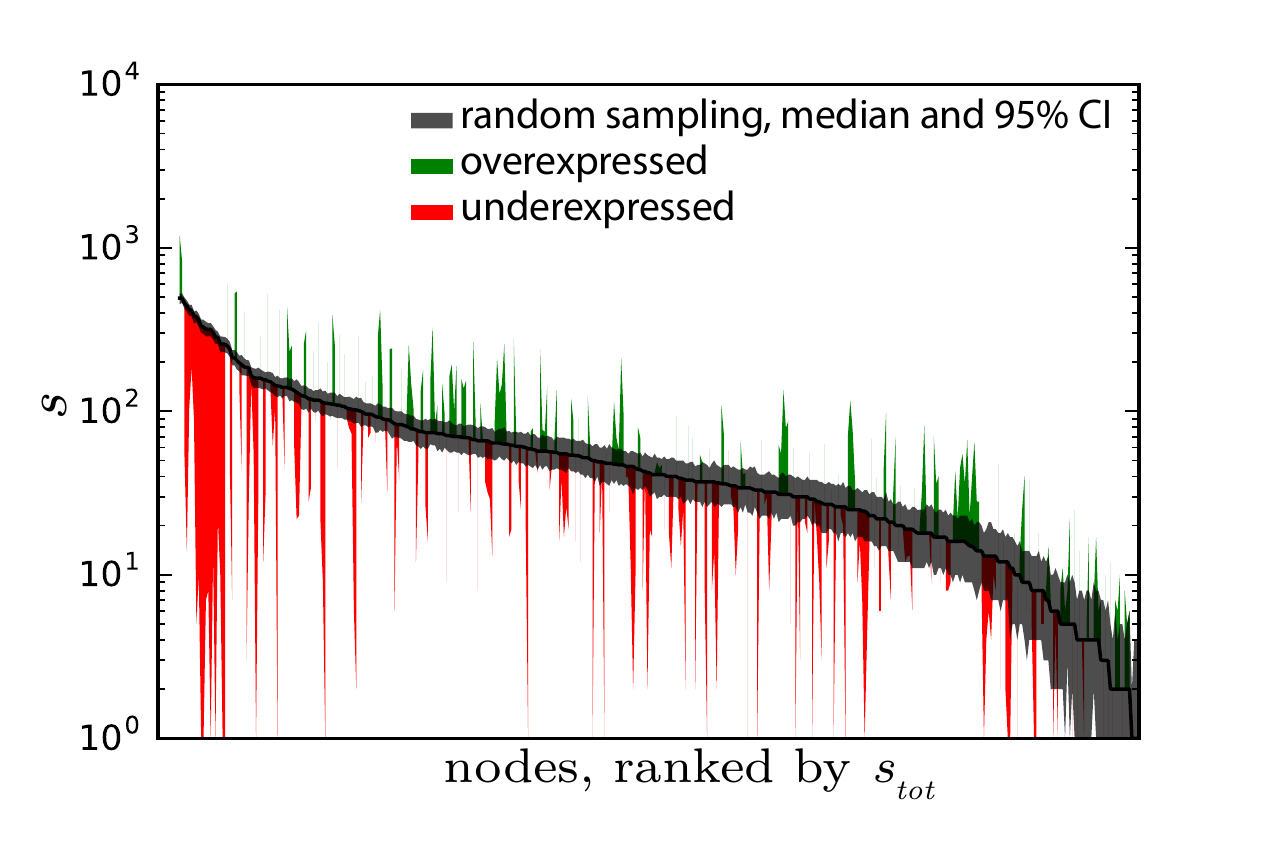} };
    \node at(-7.,1.6){\large (a)};
    \node at(4.,-0.3){ \includegraphics[width=0.55\textwidth]{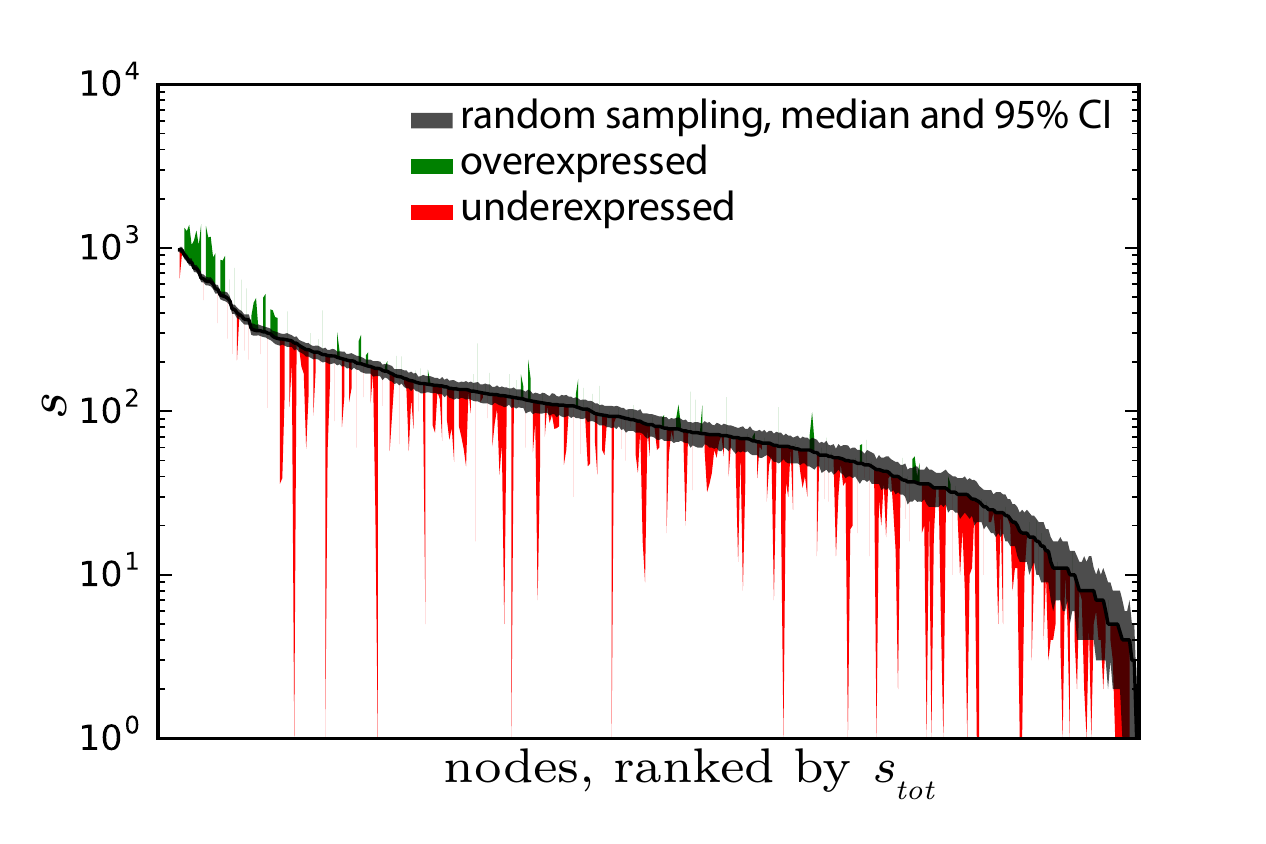} };
    \node at(0.3,1.6){\large (b)};
    \node at(-3.3,-5.){    \includegraphics[width=0.55\textwidth]{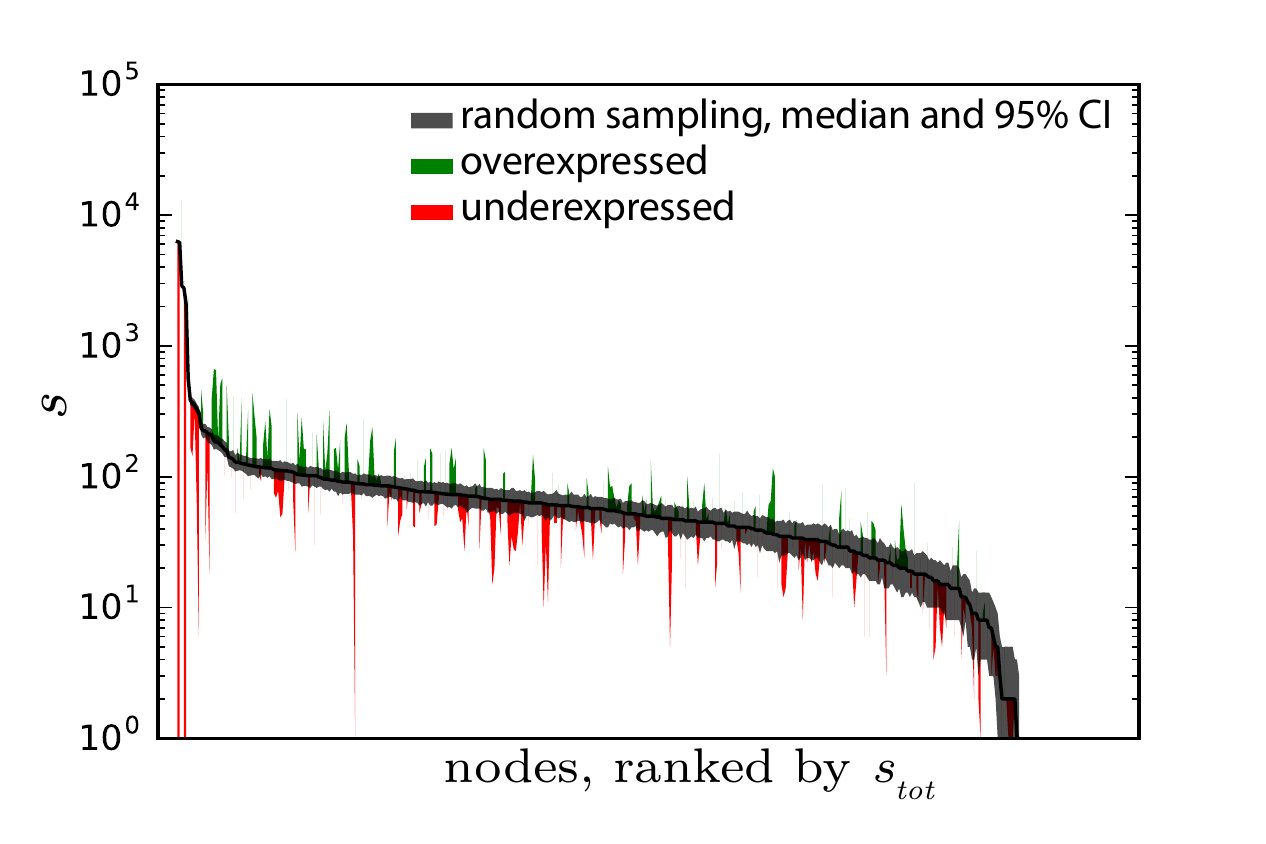} };
    \node at(-7.,-3.1){\large (c)};
    \node at(4.,-5.){ \includegraphics[width=0.55\textwidth]{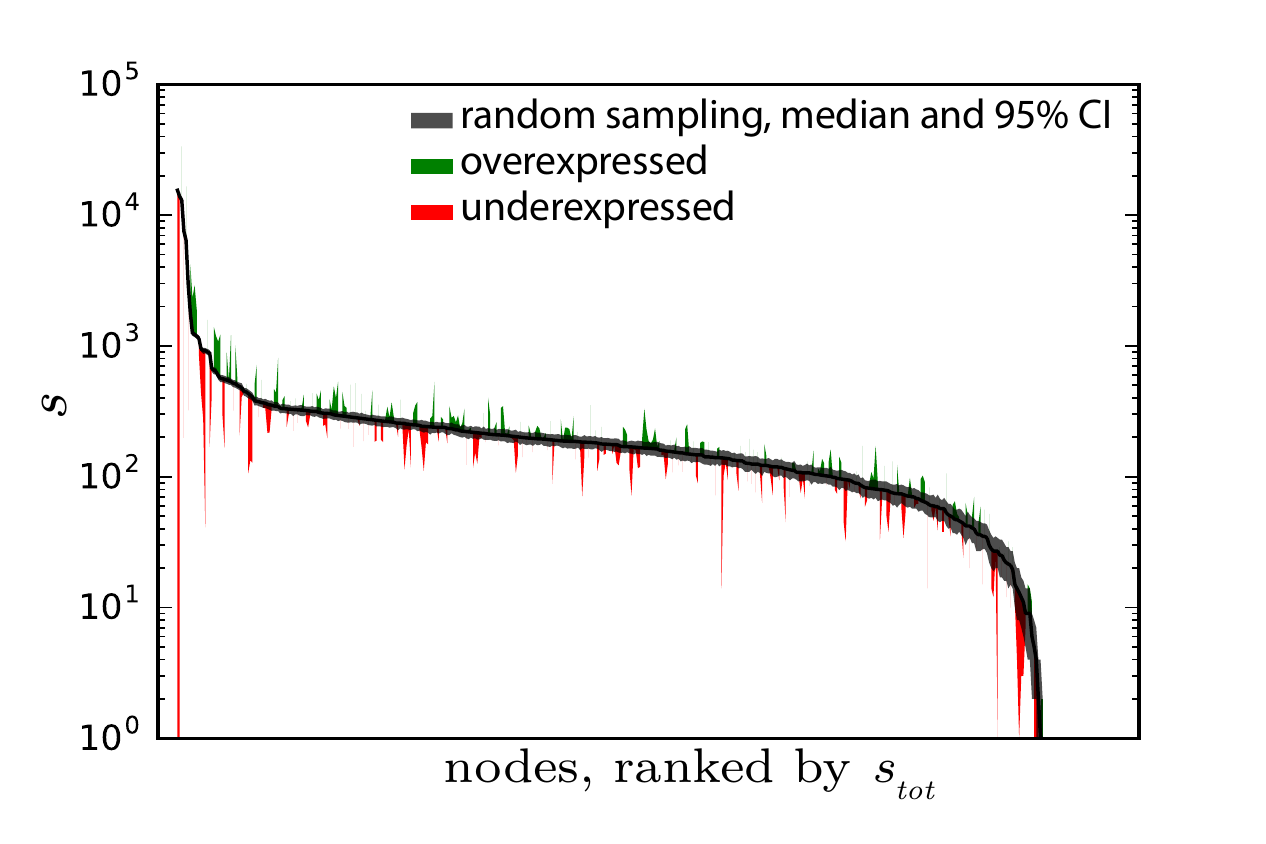} };
    \node at(0.3,-3.1){\large (d)};
  \end{tikzpicture}}
  \caption{{\bf (Color online) Contribution to the strength of nodes from the contacts occurring in a single location.}
Recorded strength $s$ (number of contacts in which a node participates) (color) in a given location,
    compared to its expected strength obtained by a random selection of the same fraction of contacts from the full network (black line:
    median; grey area shows $95\%$ C.I. from $100$ realizations). Nodes are ranked by their strength $s_{tot}$ in the full network (from highest to
    lowest).  (a), (b) Locations $3$ and $4$ of the empirical network, containing respectively $22.2\%$ and $43.7\%$ of the total contacts.
    (c),(d) locations of the synthetic model containing respectively $20\%$ and $38.8\%$ of the total contacts.  In location $3$ of the
    empirical network, top ranking nodes have lower strength than would result from random sampling. In location $4$, top ranking nodes
    have higher strength than expected. In synthetic networks, top ranking nodes are neither systematically under- nor over-expressed.
}
  \label{fig:plot_strength_behavior}
\end{figure}

\section{Conclusions}
In this paper, we have investigated, using synthetic and empirical temporal networks of human face-to-face interactions,
how spatially constrained sampling, due to partial monitoring of the various locations in which contacts can occur,
impacts the estimation of the epidemic risk in the population under study.
Such sampling leads to a systematic overestimation of the epidemic threshold, i.e., to an underestimation of the epidemic risk.
Interestingly however, this underestimation is substantially smaller than the one obtained by a random sampling of the same fraction of contacts and it becomes negligible
for high enough coverage (when the fraction of sampled contacts is higher than $\sim60\%$).
The qualitative behaviors obtained in resampled empirical and synthetic contact data are similar. However, we observe some disparities
due to the simplifying assumption of similar behavior in different locations made in the model used to produce the synthetic data, which does
not hold in real settings. If specific locations in which the hubs turn out to have most contacts are monitored, the epidemic threshold
computed on partial data can be much closer to the one obtained with the full data set than what would be expected from the
example of the synthetic data.
Further investigations with more complex models could shed more light on this issue.

The results presented here could also serve as
a starting point for the development of systematic procedures able to
produce an estimate of the real epidemic risk even when only sampled
data is available. As done recently for the case of uniform population sampling~\cite{Genois2015b}, a sensible
procedure would be to combine the known, sampled data with surrogate data describing the unknown
contacts taking place in the non-monitored location. Such surrogate data could be built as surrogate
timelines of contacts between individuals present in the non-monitored location, in a way to
respect the distributions of contact and inter-contact durations measured in the monitored one:
such distributions have indeed  be found to be very robust and are thus expected to be the same in different
locations  \cite{Cattuto2010,Isella2011,Genois2015a,Genois2015b}. An important
issue remains however open:
in contrast with the case of population sampling, one cannot easily
extrapolate the number and frequency of contacts in the non-monitored place  from
the data observed in the monitored area, as they could
correspond to very different amounts of overall contact activity.
Additional information concerning the specificities of the monitored and non-monitored locations would then be
necessary for this purpose in realistic settings. Further work will investigate how to deal with this issue and to which extent possible estimates of the epidemic threshold would depend on the assumptions made to produce surrogate data for the non-monitored location.

\section*{Acknowledgments}
The authors thank the SocioPatterns collaboration for privileged
access to the empirical data set used in this paper.
The present work was partially supported by the French ANR project HarMS-flu (ANR-12-MONU-0018) to M.G., V.C., A.B., and E.V.,
by the EU FET project Multiplex 317532 to A.B. and C.L.V., by the A*MIDEX project (ANR-11-IDEX-0001-02)
funded by the {\sl Investissements d'Avenir} French Government program and managed by the French National Research Agency (ANR) to A.B.

\bibliographystyle{naturemag}
\bibliography{dynamic_arxiv}

\clearpage
\setcounter{figure}{0}
\setcounter{page}{20}

~\vspace{3cm}~

\center{\bf\Large Supplementary Figures}
\thispagestyle{empty}

~\vspace{3cm}~

\renewcommand{\figurename}{{\bf Supplementary Figure}}

\begin{figure}
  \includegraphics[width=\textwidth]{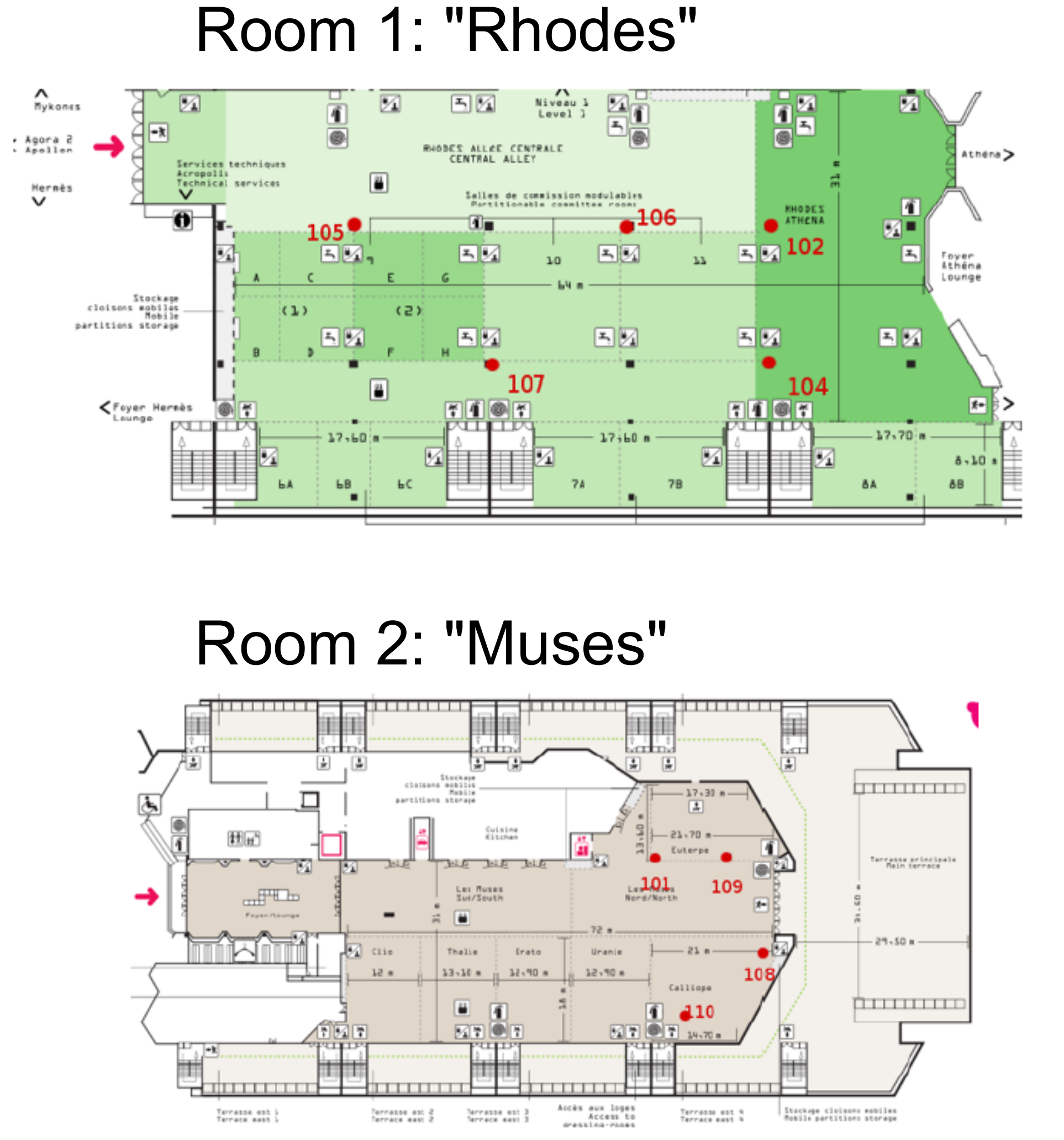}
  \caption{{\bf Map of the conference rooms and locations of receivers.} Red dots mark the location of
 the corresponding receivers. Receivers 100, 103 and 111 are located outside the two rooms and are not shown on this map.}
  \label{fig:Map}
\end{figure}

\end{document}